\documentstyle[11pt,appb]{article} 

\begin{document}

\title{MINIMAL COUPLING AND THE EQUIVALENCE PRINCIPLE 
IN QUANTUM MECHANICS\thanks{Presented at the workshop on 
Gauge Theories of Gravitation, Jadwisin, Poland, 
September 4 -- 10, 1997. \hfill\break 
Appeared in {Acta Physica Polonica {\bf 29}, 1057 (1998)}}}
\author{Claus L\"ammerzahl
\thanks{e--mail:claus@spock.physik.uni-konstanz.de}
\address{Laboratoire de Physique des Lasers, Insitut Galil\'ee \\ 
Universit\'e Paris 13, F -- 93430 Villetaneuse, France \\ 
and \\  
Fakult\"at f\"ur Physik, Universit\"at Konstanz \\ 
Fach M 674, D -- 78457 Konstanz, Germany}
}

\maketitle

\begin{abstract}
\begin{center}
{\it Dedicated to Andrzej Trautman in honour of his 64${}^{\hbox{th}}$ birthday}
\end{center}

\bigskip
The role of the Equivalence Principle (EP) in classical and quantum mechanics is reviewed. 
It is shown that the weak EP has a counterpart in quantum theory, a Quantum 
Equivalence Principle (QEP). 
This implies that also in the quantum domain the geometrisation of the gravitational interaction is an operational procedure similar to the procedure in classical physics. 
This QEP can be used for showing that it is only the usual Schr\"odinger equation coupled to gravito--inertial fields which obeys our equivalence principle. 
In addition, the QEP applied to a generalised Pauli equation including spin 
results in a characterisation of the gravitational fields which can be 
identified with the Newtonian potential and with torsion. 
Also, in the classical limit it is possible to state beside the usual EP 
for the path an EP for the spin which again may be used for introducing 
torsion as a gravitational field. 
\end{abstract}

  
\section{Introduction}

The Equivalence Principle (EP) is a statement of universality in 
the sense that a certain physical effect occurs for all 
members of a class of physical objects. 
Equivalently, the EP is a statement of independence in the sense that this physical effect occurs 
independent of the chosen member of the class of physical objects. 
The importance of an EP lies in the fact that it serves as a tool to characterise or to distinguish 
the gravitational interaction from other interactions. 
Indeed, by means of the EP we can {\it define} what we mean by a gravitational field. 

The validity of the EP is an important matter of discussion in connection with the attempt of unifying gravity with the other three interactions or with the program of quantising gravity. 
Any violation of this basic principle of General Relativity 
has of course dramatic implications for our present understanding of the 
physical nature of gravitation and of the cosmic evolution of our universe. 

According to the physical domain under consideration there may be many 
versions of EPs. 
Each EP depends on the class of chosen physical objects and for each class 
EPs of various strength may be formulated. 

\subsection{The EP in classical physics}

The {\it weak EP} (or the universality of free fall) requires that in a 
gravitational field $(i)$ {\it all} structureless point particles follow the same path and $(ii)$ this path is uniquely characterised by the initial position $x(t_0)$ and initial velocity $\dot x(t_0)$. 
This leads to an equation of motion for the path of the form  
$\ddot x^\mu = f^\mu(x, \dot x)$ where the dot means the derivative with respect to some parameter. 
Since in this equation there appears no parameter characterising properties of the particle,  gravity acts universal and can be geometrised. 
Here gravity is connected with the function $f^\mu(x, \dot x)$. 
Gravity may still be given by a Finslerian geometry, for example. 

In order that gravity is an affine geometry one has to apply the {\it strong EP} or Einstein's elevator which means that gravity can be transformed away by choosing an appropriate frame of reference (this also means that gravity can be simulated by choosing some frame of reference).  
The strong EP requires that for each point $x_0$ there is a frame so that for all particles $\ddot x = 0$ in $x_0$. 
This specifies the functions $f^\mu(x, \dot x)$ and leads to the path
equation $\ddot x^\mu + \Gamma^\mu_{\nu\rho}(x) \dot x^\nu \dot x^\rho 
\sim \dot x^\mu$ thus introducing the notion of a (projective) connection. 
Here gravity is described by the functions $\Gamma^\mu_{\nu\rho}(x)$.
(Another version of the strong EP requires that for all $x_0$ there is 
a frame so that locally all physical phenomena are as in gravity--free space. 
However, this version is not operational since one has to know in advance non--gravitational physics which is not possible because gravity cannot be shielded.)

If gravity should be describable within the framework of a Riemannian 
geometry, that is, by a space--time metric, then {\it Einstein's EP} has to be applied, \cite{ThorneLeeLightman73,Will93}. 
Einstein's EP requires that the weak EP, 
local Lorentz invariance and local position invariance holds.
Local Lorentz invariance means that gravity is described by a metric and 
possibly additional scalar fields. 
However, these scalar gravitational fields are ruled out by local 
position invariance. 

Einstein's EP implies the strong EP, and the stong EP implies the weak EP. 
Therefore we have a hierarchy of EPs, each leading to a 
more specific geometrical frame for the description of gravity. 

The main point of the above discussion is that the EP is a means $(i)$ to {\it define} operationally the gravitational interaction, $(ii)$ 
to {\it geometrise} the gravitational interaction, and $(iii)$ to fix the {\it equation of motion} of matter. 

\subsection{An EP in the quantum domain?}

Due to the very different nature of physical phenomena in the quantum regime one may ask whether there is any hope at all to find an EP in the quantum domain. 
In classical physics the EP is a strictly local notion since it involves the notions of a point or a path of a point particle. 
In the quantum domain physics is described by fields which are nonlocal objects since they are spread out over all space. 
Consequently, one has to doubt whether local notions in a formulation of an 
EP will make sense in the quantum domain. 

The EP in quantum mechanics can be discussed with respect to 
$(i)$ the minimal coupling procedure \cite{AHL92}, $(ii)$ the path (WKB--path, 
Heisenberg equations of motion \cite{Rumpf79}, path integral \cite{Kleinert78,Pelster78}), $(iii)$ the 
phase shift in neutron interferometry \cite{COW75,Greenberger68}, $(iv)$ the question 
whether gravitationally induced effects can be transformed away \cite{AHL92}, and 
$(v)$ with respect to the structure of the solution and of the Green 
function.
Therefore there are many attempts to discuss this notion or to transform this notion into the quantum domain. 
However, these approaches are $(i)$ either valid for a restricted domain only 
(classical approximation), or $(ii)$ disprove that this notion can be 
carried over to the quantum domain, or $(iii)$ are valid only for homogeneous 
gravitational fields, or $(iv)$ are rather formal statements which have no clear operational realisation. 
For example, the solution of the Schr\"odinger equation in a homogenous 
gravitational field depends on mass contrary to the analogous situation for point particles.  
Also, the phase shift in neutron interferometry in a homogeneous gravitational field depends on the mass, too, again violating the weak EP \cite{COW75,Greenberger68}. 
Also the strong EP is not valid for neutron interferometry since curvature 
effects which are present due to the extension of the interferometer, cannot be transformed away by choosing a distinguished reference frame.   

\begin{figure}
\begin{center}
{\sf\unitlength0.6cm
\begin{picture}(12,5)
\put(2.8,1.4){\makebox(0,0){A}}
\put(8.8,1.4){\makebox(0,0){B}}
\put(4,5.4){\makebox(0,0){C}}
\put(10,5.4){\makebox(0,0){D}}
\put(0,1.4){\makebox(0,0){\footnotesize neutron source}}
\put(12,5.4){\makebox(0,0){\footnotesize interference}}
\put(4.5,0.5){\vector(-1,0){2.5}}
\put(5.5,0.5){\vector(1,0){2.5}}
\put(5,0.5){\makebox(0,0){\sl l}}
\put(6,2.5){\vector(0,-1){1.5}}
\put(6,3.5){\vector(0,1){1.5}}
\put(6,3){\makebox(0,0){\sl h}}
\put(10.5,3){\makebox(0,0){$- \nabla U$}}
\put(11.5,3.5){\vector(0,-1){1}}
\thicklines{
\put(0,1){\vector(1,0){0.8}}
\put(0.8,1){\vector(1,0){0.4}}
\put(1.2,1){\line(1,0){0.8}}
\put(2,1){\vector(1,0){3}}
\put(5,1){\line(1,0){3}}
\put(2,1){\vector(1,2){1}}
\put(3,3){\line(1,2){1}}
\put(4,5){\vector(1,0){3}}
\put(7,5){\line(1,0){3}}
\put(8,1){\vector(1,2){1}}
\put(9,3){\line(1,2){1}}
\put(10,5){\vector(1,0){0.8}}
\put(10.8,5){\vector(1,0){0.4}}
\put(11.2,5){\line(1,0){0.8}}}
\end{picture}}
\end{center}
\caption{Neutron interference used in \cite{COW75} 
to demonstrate the 
gravitational influence on the neutron's wave function. $A$ and $D$ are 
beam splitter and recombiner, respectively, and $B$ and $C$ are 
mirrors. \label{Fig:COW}}
\end{figure}
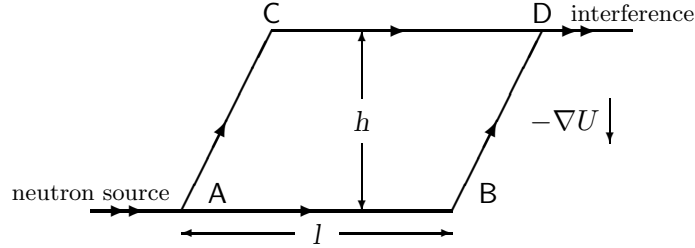

To be more explicit, the phase shift in neutron interferometry (see Fig.\ref{Fig:COW}) is described using the Schr\"odinger equation coupled to the Newtonian potential 
and performing a WKB approximation \cite{COW75,La96}. 
The measured quantity is the intensity of the neutron beam leaving the 
interferometer at one port, 
$I = \frac{1}{2} I_0 (1 - \cos\phi_{\hbox{\scriptsize grav}}^{\hbox{\scriptsize neutron}})$, with the phase 
\begin{equation}
\phi_{\hbox{\scriptsize grav}}^{\hbox{\scriptsize 
neutron}} = \frac{1}{\hbar}\oint \mbox{\boldmath$p$}
(\mbox{\boldmath$x$}) \cdot d \mbox{\boldmath$x$} = \frac{m \mbox{\boldmath$g$}\cdot\mbox{\boldmath$h$} l}
{\hbar v} + U_{ab} h^a h^b \frac{m l}{\hbar v} + 
{\cal O}(g^2, g U_{ab})   \label{NeutronPhase}
\end{equation}
($\mbox{\boldmath$g$} := \nabla U$). 
The phase shift depends on the mass $m$ of the neutron so that there is clearly no weak EP, and it depends on the  
curvature $U_{ab} := \partial_a \partial_b U$ so that also the strong EP is not valid.
We want to stress the most characteristic feature of this derivation of 
the phase shift: it treats the beam splitting as a splitting in 
{\it configuration space}. 

\subsection{General remarks}

In order to geometrise gravity 
it is enough to have {\it one effect} for which an EP is valid. 
This is important because there are of course effects in classical and quantum mechanics which depend on the mass of the considered object. 
For example, the elastic scattering of point particles or the spacing between the interference fringes in a double slit interference experiment. 
Therefore all EPs are valid for distinguished situations only which may be 
very difficult to specify experimentally.   

The importance of the problem whether there is still some EP in the 
quantum domain concerns the notion of the gravitational field and the 
space--time geometry in the quantum domain: if and only if there is an EP 
then also in the quantum domain one can define uniquely and without any 
approximation what we mean by the gravitational field in this domain. 
Only in this case we are allowed to geometrise gravity in the domain 
of quantum physics.
We will show in the following that there is an EP in 
quantum theory which is slightly modified compared to the weak EP 
in classical mechanics. 
However, it has the same logical structure than all other EPs.
Therefore, {\it also in the quantum domain gravitation is 
operationally definable and geometrisable}. Here we will also include 
spin in our considerations.  

There are two points concerning the search for an EP within a certain 
theory: On the one hand, if a certain theory is given then we may ask which EP is valid for this theory. 
On the other hand we may also ask which theory is singled out of a wide range of theories by the required validity of an EP. 
That is what we are doing in Scts.\ref{Sctn:GenSchr} and \ref{Sctn:GenPauli}.

Another point is the following: In the quantum domain one usually employes a minimal coupling procedure in order couple gravity to the Klein--Gordon or the Dirac equation. 
However, in doing so one has to know in advance the physics in the gravity--free world. 
Since there is always a gravitational field all over the universe, this is neither an operational nor a self--contained approach. 
However, in the following we are able to show that at least in the 
non--relativitistic domain for first quantised matter {\it the minimal 
coupling procedure is equivalent to the QEP} which we are going to formulate.  

\section{A Quantum Equivalence Principle}

In order to give motivation for our formulation of a Quantum Equivalence Principle (QEP) we first discuss the phase shift for atom beam interferometry in gravitational fields. 
This structure of the corresponding result allows us to state a QEP. 
In addition, we are able to show that this QEP can also be applied to neutron interferometry. 
For more discussions, see \cite{La96}. 

\subsection{Atom beam interferometry in gravitational fields}

\begin{figure}
\begin{center}
\unitlength0.54cm
\qquad\qquad\begin{picture}(12.5,8)
\put(0,0.5){\vector(1,0){12}}
\put(0,0.5){\vector(0,1){7}}
\put(0.5,1.5){\vector(1,0){1.5}}
\put(2,1.5){\vector(1,4){0.5}}
\put(2.5,3.5){\vector(1,0){3}}
\put(2,1.5){\vector(1,0){3.5}}
\put(5.5,3.5){\vector(1,-4){0.5}}
\put(5.5,1.5){\vector(1,4){0.5}}
\put(6,1.5){\vector(1,0){3}}
\put(6,3.5){\vector(1,0){3}}
\put(9,3.5){\vector(1,-4){0.5}}
\put(9,1.5){\vector(1,4){0.5}}
\put(9,3.5){\vector(1,0){0.5}}
\put(9,1.5){\vector(1,0){0.5}}
\put(9.5,1.5){\vector(1,0){2}}
\put(9.5,3.5){\vector(1,0){2}}
\put(2.1,6.5){\vector(0,-1){1.8}}
\put(2.4,6.5){\vector(0,-1){1.8}}
\put(5.45,6.5){\vector(0,-1){1.8}}
\put(5.75,6.5){\vector(0,-1){1.8}}
\put(6.05,6.5){\vector(0,-1){1.8}}
\put(9.1,6.5){\vector(0,-1){1.8}}
\put(9.4,6.5){\vector(0,-1){1.8}}
\put(2.5,7){\makebox(0,0){$\pi/2$--pulse}}
\put(5.75,7){\makebox(0,0){$\pi$--pulse}}
\put(9.25,7){\makebox(0,0){$\pi/2$--pulse}}
\put(3,5.5){\makebox(0,0){$k$}}
\put(-0.5,7){\makebox(0,0){$p$}}
\put(11.5,0.1){\makebox(0,0){$t$}}
\put(2.25,0){\makebox(0,0){$t_1$}}
\put(5.75,0){\makebox(0,0){$t_2$}}
\put(9.5,0){\makebox(0,0){$t_3$}}
\put(-0.3,1.4){\makebox(0,0){$p$}}
\put(-0.9,3.5){\makebox(0,0){$p + \hbar k$}}
\put(1,1.9){\makebox(0,0){$\scriptstyle | g, p\rangle$}} 
\put(4,1.9){\makebox(0,0){$\scriptstyle | g, p\rangle$}} 
\put(7.5,1.9){\makebox(0,0){$\scriptstyle | g, p\rangle$}} 
\put(10.75,1.9){\makebox(0,0){$\scriptstyle | g, p\rangle$}} 
\put(4,3.9){\makebox(0,0){$\scriptstyle | e, p + \hbar k\rangle$}} 
\put(7.5,3.9){\makebox(0,0){$\scriptstyle | e, p + \hbar k\rangle$}} 
\put(10.75,3.9){\makebox(0,0){$\scriptstyle | e, p + \hbar k\rangle$}}
\put(12,1.5){\makebox(0,0){II}}
\put(12,3.5){\makebox(0,0){I}}
\end{picture}
\vspace*{1mm}
\end{center} 
\caption{Atom beam interferometry as closed loop in momentum space. 
$| g, p\rangle$ is an atom in the ground state with momentum $p$, 
$| e, p\rangle$ is an atom in the excited state. 
The laser pulses change  the internal state as well as the momentum of the atoms thus acting as ``mirrors'' in momentum space. 
For the Ramsey--Bord\'e--interferometer we just have to replace $t$ by $x$. 
\label{Fig:AI}} 
\end{figure}
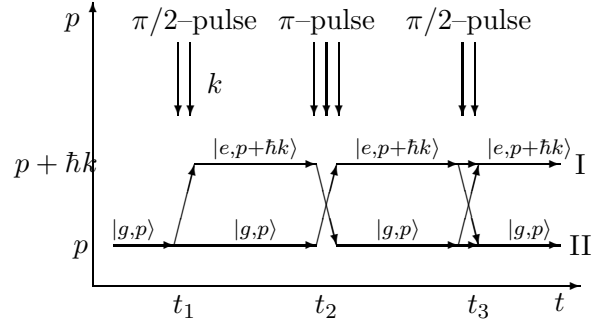

The important point is that in atom beam interferometry 
the interference is described by a closed loop in {\it momentum 
space} and not in configuration space: it is the momentum space where the 
splitting occurs (see Fig.\ref{Fig:AI}). 
The observed quantity is the number of atoms in the excited state leaving the apparatus at port I, for example: 
\begin{equation}
I_2 = \int |a_{e, \mbox{\boldmath$\scriptstyle p$} + 
\hbar \mbox{\boldmath$\scriptstyle k$}}(t_2)| d^3p = 
\frac{1}{2} \left(1 - \cos\phi\right) I_1 \, .
\end{equation}
From the dynamics of the quantum field we can calculate the phase 
shift due to the gravitational interaction \cite{La96}:
\begin{equation}
\phi_{\hbox{\scriptsize grav}}^{\hbox{\scriptsize atom}}  = - k^a T^2 \left(g_a - U_{ab} T\left(\frac{\hbar k^b}{2 m} + 
\langle \widehat v^b \rangle_0 - g^b 
\frac{37}{12} T\right)\right) \, .  \label{phase}
\end{equation} 
Here $\langle \widehat v^b \rangle_0$ is the mean value of the velocity operator at the moment of the first laser pulse.

Eqn.\ (\ref{phase}) is an exact quantum result, we made no 
approximation as for example in $\hbar$ as in the description of the neutron interferometer. (\ref{phase}) is also exact in $\mbox{\boldmath$g$}$ and of first order in $U_{ab}$. The first term $\phi_{\hbox{\scriptsize grav}} = - 
\mbox{\boldmath$k$} \cdot \mbox{\boldmath$g$}\; T^2$ 
describes the phase shift due to the acceleration. 
An astonishing feature is, that although there appears no $\hbar$, this term is an exact quantum result. 
The reason for that is that we use only the quantities $T$ and $\mbox{\boldmath$k$}$ which are given by the experimental setup. 
Only if we {\it formally} introduce classical notions then $\hbar$ as 
well as the mass $m$ comes in:   
defining formally a ``length'' by means of $\mbox{\boldmath$l$} = \langle \widehat{\mbox{\boldmath$v$}} \rangle_0 T$ and a ``height'' by 
$\displaystyle \mbox{\boldmath$h$} = (\hbar \mbox{\boldmath$k$}/m) T$ then the above acceleration induced phase shift can be rewritten as 
$\displaystyle \phi_{\hbox{\scriptsize grav}} = - 
m \mbox{\boldmath$h$} \cdot \mbox{\boldmath$g$} l/(\hbar 
|\langle\widehat{\mbox{\boldmath$v$}} \rangle_0|)$ which has exactly the form of the first term of the phase shift (\ref{NeutronPhase}) \cite{COW75}. 
However, these notions of a ``length'' and of a ``height'' have no 
operational meaning in atom interferometry since there is no 
beam splitting in configuration space (for an atomic beam with a width of 
ca.\ 1 cm the ``splitting'' may be of about 10 $\mu$m). 
For more discussions of (\ref{phase}) see \cite{La96}. 

The {\it weak EP is exactly fulfilled for this acceleration induced phase shift} since there appears no mass. 
This also means that a mass dependence of the equations of motion 
or of the solutions is no indication for a break-down of the EP on the level of observations.

The part $\phi_{\hbox{\scriptsize grav}} = 
U_{ab} k^a \langle \widehat v^b\rangle_0 T^3$ describes the phase shift 
due to the Newtonian part of the Riemannian curvature.   
It is the quantum version of the geodesic deviation equation and gives 
a genuine quantum measurement of the space--time curvature.

Since the total phase shift (\ref{phase}) depends on $m$ the weak EP is not valid. 
However, if we consider a phase shift for an infinitesimal loop in momentum 
space, that is for $\mbox{\boldmath$k$} \rightarrow 0$, and normaise it by 
dividing through its modulus $k$, then the weak EP turns out to be valid: 
\begin{equation}
\lim_{k\rightarrow 0} \frac{1}{k} \phi_{\hbox{\scriptsize grav}} = 
- \bar k^a T^2 \left(g_a + U_{ab} T\left(\langle \widehat v^b \rangle_0 - g^b \frac{37}{12} T\right)\right) 
\end{equation}
($\bar{\mbox{\boldmath$k$}} = \mbox{\boldmath$k$}/k$). 
This expression is valid in an arbitrarily curved space-time. 

Therefore also in interference experiments the influence of the Newtonian potential on the interference fringes does not depend on the used quantum matter. 
This means that with the help of the above procedure we have an 
operational means to distinguish the gravitational interaction in the quantum domain in the same way as the usual EP does in the classical domain, and thus to assign the gravitational interaction a geometrical nature. 

We use this result to proceed in formulating an EP for the quantum domain. 

\subsection{A Quantum Equivalence Principle}

\noindent {\bf Quantum Equivalence Principle (QEP):} {\it For all given 
initial states the input independent result of a physical 
experiment is independent of the characteristic parameters 
(like mass, charge) of the quantum system.}

\medskip
Here we mean by ``input'' the characterisation of the experimental 
preparation or of the influence on physical system. 
In our case it is the wave vector $\mbox{\boldmath$k$}$ which influences 
the motion of the atoms. 
``Input independent'' means that the influence given by $k$ 
approaches zero, but the resulting expression is to be normalised 
by $k$ thus giving a finite expression. 
 
Note that for this EP we do not use any local notion. We are using 
only the notions ``quantum systems'' and ``influence''.  
As the EP in classical physics, this EP is a statement of 
universality for observations thus leading to an operational 
definition of the gravitational field and its geometrisation on 
the quantum level.
Our QEP is a generalisation of the weak EP; the strong EP is not 
applicable to quantum mechanics. 
It is clear that for charged particles in an electromagnetic field the QEP 
is not valid, see \cite{La96}. 

\subsection{The Quantum EP in neutron interferometry}

After having shown that the weak EP is valid for atom beam interferometry, 
we have to reconsider the phase shift in neutron interferometry. 
It has been shown \cite{La96} that the corresponding phase shift (\ref{NeutronPhase})
can be reformulated by
\begin{equation}
\phi_{\hbox{\scriptsize grav}}^{\hbox{\scriptsize neutron}} = 
\mbox{\boldmath$g$}\cdot\mbox{\boldmath$G$}\; 
T T^\prime  + {\cal O}(g^2, U_{ab})
\end{equation}
where $\mbox{\boldmath$G$}$ is the reciprocal lattice vector of the mono--crystal 
on the Ewald sphere near the momentum of the incoming neutron. 

Since the calculation for deriving this formula is similar to that for the derivation of the phase shift in atom interferometry, the above result is the exact phase shift in the same way as (\ref{phase}) is. 
Indeed, it has been remarked by Bord\'e \cite{Borde97} that neutron 
scattering at crystals has a mathematical structure which is similar to 
that of the scattering of atoms by laser beams. 
The difference to the usual WKB treatment is, that here we do not consider neutron interferometry as a loop in configuration space, but instead as loop in momentum space. 

Therefore, for the phase shift in neutron interferometry the same remarks 
concerning the validity of the EP hold: The QEP is valid also in neutron 
interferometry, and especially for homogeneous gravitational fields the usual weak EP is valid. 
This is in contrast to \cite{Greenberger68,Werneretal80}. 

\section{Equivalence principle and minimal coupling}

After having shown that the usual Schr\"odinger equation coupled in the 
usual way to the Newtonian potential obeys our QEP, we now want to proceed 
the other way around: 
We ask for that quantum equation of motion which results from the 
requirement of validity of our QEP. 
In order to answer that question we first start from a very general structure of a Schr\"odinger equation for a scalar field. 
In this case the minimal coupling procedure is equivalent to the validity of our QEP. 
Next we use a very general approach starting from basic principles leading to a generalised Dirac and Pauli equation. 
In this case we have to discuss the role of the spin in the formulation of our QEP. 
As a main result it follows that there is a QEP so that beside the Newtonian potential torsion naturally comes out as a geometrical field in an operational way. 

\subsection{A general Schr\"odinger equation \label{Sctn:GenSchr}}

We start from an ansatz describing a general Schr\"odinger equation
\begin{equation}
H = \frac{1}{2m}\widehat{\mbox{\boldmath$p$}}^2 + V(\widehat{\mbox{\boldmath$x$}}) + 
\frac{1}{2}(\mbox{\boldmath$V$}(\widehat{\mbox{\boldmath$x$}}) \cdot \widehat{\mbox{\boldmath$p$}} + \widehat{\mbox{\boldmath$p$}}\cdot 
\mbox{\boldmath$V$}(\mbox{\boldmath$x$})) + \frac{1}{2} W^{ab} \widehat p_a \widehat p_b \label{genSchr}
\end{equation}
For the sake of simplicity we restrict our equation to be of second order only. 
In this ansatz the fields $V(\mbox{\boldmath$x$})$, $\mbox{\boldmath$V$}(\mbox{\boldmath$x$})$, 
and $W^{ab}$ are to be determined with the help of our QEP.
We also assume that $W^{ab}$ does not depend on the position and that 
$\mbox{\boldmath$V$}(\mbox{\boldmath$x$})$ is linear in the position. 
Neither these assumptions nor the restriction to a second order equation 
influence our way of reasoning. Our scheme can be carried through in full generality. 

From this general Schr\"odinger equation we can calculate in the same way as above the input independent phase shift.   
Then we {\it require that the QEP holds}.  
This implies \cite{La96} that $V/m$, $\mbox{\boldmath$V$}$, and $m W^{ab}$ must be independent of the mass $m$ 
appearing in the kinetic term of the general Schr\"odinger equation (\ref{genSchr}). 
It is convenient to introduce new functions $U := V/m$, 
$U^{ab} := m W^{ab}$ which, due to our QEP, are independent of $m$ and thus independent of the chosen quantum system. 
Insertion of these new functions into (\ref{genSchr}) gives
\begin{equation} 
H = \frac{1}{2m}(\delta^{ab} + U^{ab}) \widehat p_a \widehat p_b + 
m U(\widehat{\mbox{\boldmath$x$}}) + 
\frac{1}{2}(\mbox{\boldmath$V$}(\widehat{\mbox{\boldmath$x$}}) \cdot 
\widehat{\mbox{\boldmath$p$}} + \widehat{\mbox{\boldmath$p$}} \cdot 
\mbox{\boldmath$V$}(\widehat{\mbox{\boldmath$x$}}))   \label{QEPSchroe}
\end{equation}
The expression $\delta^{ab} + U^{ab}$ can be 
transformed to $\delta^{ab}$ by means of a coordinate transformation.
Clearly, $U$ has to identified with the Newtonian potential. 
The QEP forces the $m$ in front of the Newtonian potential to be the same 
as in the kinetic term. 
The term $\frac{1}{2}(\mbox{\boldmath$V$}(\widehat{\mbox{\boldmath$x$}}) \cdot 
\widehat{\mbox{\boldmath$p$}} + \widehat{\mbox{\boldmath$p$}} 
\cdot \mbox{\boldmath$V$}(\widehat{\mbox{\boldmath$x$}})) = \Lambda^a_b 
\frac{1}{2} \{\widehat p_a, \widehat x^b\}$ describes an 
expanding, rotating or shearing frame.   
For an antisymmetric $\Lambda^a_b$, for example, we get the 
Sagnac--effect $\phi_{\hbox{\scriptsize Sagnac}} = 2 T^2 k^a \Lambda^b_a 
\langle \widehat v_b\rangle_0 = 2 T^2 \mbox{\boldmath$k$}\cdot(\mbox{\boldmath$\omega$} \times \mbox{\boldmath$v$})$. 
A formal us ef a ``length'' and ``height'' as above, again leads to the usual 
formula for the Sagnac effect $2 \frac{m}{\hbar} 
\mbox{\boldmath$\omega$}\cdot\mbox{\boldmath$A$}$ which contains 
$\hbar$ and the mass $m$. 

Therefore we have the result that for a general Schr\"odinger equation 
the requirement, that the QEP should hold, 
implies the usual structure of the gravito--inertial interaction in 
the quantum domain. 
The QEP amounts to an operational justification of the minimal 
coupling procedure at least in the non-relativistic domain. 
A corresponding treatment of the relativistic case should also 
be done. 

\subsection{A general approach with spin \label{Sctn:GenPauli}}

In this approach we take the quantum field as fundamental physical object 
and require dynamical principles for this quantum field which is described 
by means of a multicomponent complex valued field. 
We require the following dynamical properties:
$(i)$ there should be a well posed Cauchy--problem, 
$(ii)$ the superposition principle should hold, 
$(iii)$ it should propagate with a finite speed, 
and $(iv)$ should obey a conservation law. 
The mathematical consequence of these requirements is a generalised Dirac equation (for a review see \cite{AL92})
\begin{equation} 
0 = i \widetilde\gamma^\mu(x) \partial_\mu \varphi(x) - 
M(x)\varphi(x)
\end{equation}
which is a first order hyperbolic system of partial differential equations with still unspecified matrices $\widetilde\gamma^\mu$ and $M$.
In general, the $\widetilde\gamma^\mu$ do not fulfill a Clifford algebra. 
One can introduce the deviation $X^{\mu\nu}$ from the usual Clifford algebra $\widetilde\gamma^\mu \widetilde\gamma^\nu + 
\widetilde\gamma^\nu \widetilde\gamma^\mu = 
g^{\mu\nu} \hbox{\bf 1} + X^{\mu\nu}$ where $g^{\mu\nu} := \frac{1}{4} \hbox{tr} (\widetilde\gamma^\mu \widetilde\gamma^\nu)$. 
However, it can be shown that the notion of a generalised Clifford 
algebra always exists \cite{La98a}.
This deviation from the usual Clifford algebra can be geometrically interpreted: 
The null cones of the generalised Dirac equation which are derived from the characteristic surfaces, split in more than one components thus leading to a 
birefringence behaviour of null propagation. 
Similarly, there is also no longer a single mass shell \cite{La98a,La98}. 

In the non--relativistic limit we get a 
{\it generalised Pauli--equation} in a non--rotating frame (see \cite{La98} for the details)
\begin{eqnarray}
i \hbar \frac{\partial}{\partial t} \varphi & = & - \frac{\hbar^2}{2m} 
\left(\delta^{ij} - \frac{\delta m_{\hbox{\scriptsize i}}^{ij}}{m} - 
\frac{\delta \bar m_{\hbox{\scriptsize i} k}^{ij} \sigma^k}{m}\right)
 \nabla_i \nabla_j \varphi 
+ \left(c A^i_j + \frac{1}{m} a^i_j\right) \sigma^j i \hbar \nabla_i 
\varphi \nonumber\\
& & + \Bigl[e \phi(x) + \frac{e}{2 m} 
\mbox{\boldmath$H$}\cdot\mbox{\boldmath$\sigma$} +  
(1 + \mbox{\boldmath$C$}\cdot \mbox{\boldmath$\sigma$}) m U(x) +
\delta m_{\hbox{\scriptsize g}ij} U^{ij}(x) \nonumber\\
& & + \hbar c \mbox{\boldmath$T$}\cdot
\mbox{\boldmath$\sigma$} + 
m c^2 \mbox{\boldmath$B$}\cdot\mbox{\boldmath$\sigma$}\Bigr] \varphi \label{GPE}
\end{eqnarray} 
In this equation there appear several anomalous terms which are not 
present in the usual Pauli equation coupled to the Newtonian potential: 
the terms $\delta m_{\hbox{\scriptsize i}}^{ij}$ and 
$\delta \bar m_{\hbox{\scriptsize i} k}^{ij}$ describe an anomalous inertial 
mass which may depend on the spin of the particle, 
$A^i_j$ and $a^i_j$ give rise to a spin--momentum coupling, 
$\mbox{\boldmath$C$}$ characterises an anomalous spin coupling to the 
Newtonian potential, $\delta m_{\hbox{\scriptsize g}ij}$ is the anomalous 
gravitational mass tensor, $\mbox{\boldmath$T$}$ represents an extra spin--coupling term, 
and $\mbox{\boldmath$B$}$ may be called 
a spin--dependent ``rest mass''. 
These anomalous terms violate Einstein's EP \cite{La98}. 
$\phi$ and $\mbox{\boldmath$H$}$ are the scalar electrostatic potential and the magnetic field, respectively. 

\paragraph{Interference experiments}

We describe two types of interference experiments which are useful for our 
attempt to apply the QEP. 
For a spin--flip interference experiment we use an atomic beam in a defined spin state, say $|\frac{1}{2}, \frac{1}{2}\rangle$. 
A beam splitter splits via a spin--flip this state into a superposition of the states $|\frac{1}{2}, \frac{1}{2}\rangle$ and $|\frac{1}{2}, -\frac{1}{2}\rangle$. 
After a time $\Delta t$ again a spin--flip is applied recombining these two states. 
Measuring again the spin along the given axis, gives an interference 
pattern depending on the difference of the energies accumulated by these two states during the time $\Delta t$. 
The corresponding phase is given by 
\begin{equation}
\phi^{\hbox{\scriptsize spin}} = {2\over\hbar} \left({{\delta\bar
m_{\hbox{\scriptsize i} k}^{ij}}\over{2 m^2}} p_i p_j -
{\hbar \over m} a^i_k p_i -
c A^i_k p_i + m c^2 B_k + C_k m U + c T_k\right) S^k \Delta t
 \label{spinphase} 
\end{equation}

A second type of experiment measures the acceleration, compare \cite{KasevichChu}. 
The corresponding phase shift for a spherically symmetric gravitational field disregarding curvature is
\begin{eqnarray}
\phi^{\hbox{\scriptsize accel}}& = & T^2
{{GM}\over{r_0^3}}\left(k_i r^i_0 +
{{\delta m_{\hbox{\scriptsize g} ij} }\over m} {{r_0^i
r_0^j}\over{r_0^2}}
k_l r^l_0 - {{6}\over
5}{{\delta m_{\hbox{\scriptsize g} ij}}\over m} r_0^i k^j \right. \nonumber\\
& & \left.  +
{2\over 5} {{\delta m_{\hbox{\scriptsize g} ii}}\over m}
k_l r^l_0 - {{\delta m_{\hbox{\scriptsize i} ij} + \delta \bar m_{\hbox{\scriptsize i} ijk} S^k }\over m}
r_0^i k^j + C_j S^j k_i r^i_0 \right) \, ,
 \label{accelphase}
\end{eqnarray}
where $S^i$ is the spin of the quantum particle.

Before applying our QEP we present the equations of motion of the classical limit of 
the generalised Pauli equation. 

\paragraph{The classical limit}

From eqn.\ (\ref{GPE}) we get as classical acceleration
\begin{equation}
a^i =- \left(\delta^{ij} + 
\frac{\delta m_{\hbox{\scriptsize i}}^{ij}}{m} + 
\left(\frac{\delta\bar m_{\hbox{\scriptsize i} k}^{ij}}{m} + 
\delta^{ij} C_k\right) S^k\right) \partial_j U - 
\delta^{ij} \frac{\delta m_{\hbox{\scriptsize g} kl}}{m} 
\partial_j U^{kl} 
\end{equation}
We recognise that on the quantum level (\ref{spinphase},\ref{accelphase}) there are more 
possibilities to violate the weak EP than on the classical level. 

We also can calculate the dynamical behavior of the spin expectation value in the classical approximation:
\begin{equation}  \label{SpinMotion}
\frac{d}{dt} \mbox{\boldmath$S$} = \mbox{\boldmath$\Omega$} \times \mbox{\boldmath$S$}
\end{equation}
with
\begin{equation}
\Omega_i := \frac{1}{2 m} \frac{\delta \bar m_{\hbox{\scriptsize i} i}^{kl}}{m} p_k p_l + 
\left(\frac{1}{m} a^j_i + c A^j_i\right) p_i + m c^2 B_i + c\, T_i + 
C_im U(x)
\end{equation}
In order to test the complete set of anomalous parameters we need the 
equation of motion for the path {\it and} of the spin. 
Both quantities are needed in order to determine the structure of space--time. 
This point of view is the basis of Riemann--Cartan theories as has been 
stressed e.g.\ by Trautman \cite{Trautman66} and Hehl \cite{Hehletal76}. 

\subsection{Equivalence principles}

Based on our QEP we can now state two different EPs. This is possible 
because our quantum system now has two properties: mass and spin. 
One QEP requires that the measured effects are independent of the mass, 
and the second requires the independence from mass {\it and} spin.  

\medskip
\noindent ${\hbox{\bf QEP}}_m$: The input independent phase shift should be independent of the mass of the particle.

\medskip
If we apply ${\hbox{QEP}}_{\! m}$ to the acceleration induced phase shift $\phi^{\hbox{\scriptsize accel}}$ then we get as necessary conditions $\delta m_{\hbox{\scriptsize i}}^{kl}$, $\delta \bar m_{\hbox{\scriptsize i} i}^{kl}$, $C_k$, $\delta m_{\hbox{\scriptsize g} ij} = 0$. 
${\hbox{QEP}}_m$ applied to the spin--flip induced phase shift $\phi^{\hbox{\scriptsize spin}}$ gives $B_i$, $A^i_j$, $\delta \bar m_{\hbox{\scriptsize i} i}^{kl} = 0$. 
Taking both results together we have that all anomalous terms but $\mbox{\boldmath$T$}$ and $a^i_j$ have to vanish. 
This is also clear from the fact that the coupling to $\mbox{\boldmath$T$}$ 
and $a^i_j$ in (\ref{GPE}) are the only one which do not involve the mass of the quantum object and thus 
can be regarded as coupling to a geometrical field. 
Indeed, it is possible to absorb the $a^i_j$--term into the kinetic term by replacing $\nabla_i \rightarrow \nabla_i - \frac{i}{\hbar} \delta_{ij} a^j_l \sigma^l$ and neglecting terms of second order in the anomalous terms. 
According to the non--relativistic limit of the Dirac equation 
in a Riemann--Cartan space--time \cite{La97}, these terms can be 
interpreted as time-- and space--components of the axial torsion. 
Since the Dirac equation in Riemann--Cartan space--time was obtainad by minimally coupling of the corresponding Lagrangian (see e.g.\ \cite{Hehletal76}), our QEP again singles out this procedure, at least in the non--relativitic domain considered here. 

As a second possibility we can regard spin to be on the same level as mass being a property of quantum objects. 
Then we can take as our equivalence principle the requirement that the input independent phase shift should be independent of the mass and the spin of the particle. 
This we call ${\hbox{QEP}}_{m, S}$. 

\medskip
\noindent ${\hbox{\bf QEP}}_{\! m,S}$: The input independent phase shift 
should be independent of the mass {\it and} the spin of the particle.

\medskip
Then we get the same results as above but in addition that $\mbox{\boldmath$T$}$ and $a^i_j$ have to vanish, too. 

Therefore, if we regard spin on the same level as mass, then the requirement of a QEP rules out torsion. 
However, if we regard mass to be more fundamental in this connection, then 
torsion is still allowed as a geometrical field interacting with quantum objects. 
Therefore, there is a QEP, namely ${\hbox{QEP}}_{\! m}$, so that the Newtonian potential (or the space--time metric) as well as torsion can be introduced operationally as geometric fields. 
The requirement that the QEP should hold for the generalised Pauli equation 
implies that all couplings except the Newtonian potential and the axial 
torsion should vanish. 
Consequently, {\it at least in the non--relativistic regime, Riemann--Cartan geometry is a consequence of a QEP}.

There is another possibility to formulate an EP, namely on the level of the 
equations of motion for the path and for the spin in the quasiclassical approximation. 
The structure of the equation of motion for the spin (\ref{SpinMotion}) suggests an EP for the spin which also naturally leads to the introduction of torsion as the only geometrically interpretable interaction with the spin. 

\medskip
\noindent {\bf EP for the spin:} For any initial state of the quantum system, 
the spin motion does not depend on the characteristic parameters of the 
quantum system (like mass, charge, etc.). 

\medskip
A similar EP for the spin has been first suggested by Adamowicz and Trautman \cite{AdamowiczTrautmann75}.
The only terms in (\ref{SpinMotion}) which do not depend on the mass or on the 
charge of the particle are the terms $a^i_j$ and $\mbox{\boldmath$T$}$. 
Since these terms can be interpreted as parts of axial torsion, 
the requirement of the EP for the spin implies that all interactions except the Newtonian potential and the axial torsion should vanish leading again to a Riemann--Cartan geometry.

Also within a constructive axiomatic approach it was possible to introduce torsion from the equation of motion for the spin \cite{AudretschLa88}. 
Of course, torsion violates local Lorentz invariance so that Einstein's EP is not valid. 

\section{Conclusion}

We have shown that also in the quantum domain it is possible to 
state an EP which distinguishes the gravitational fields from other 
fields. 
Therefore, also in the quantum domain gravity is geometrisable and the 
notion of a space--time structure makes operational sense in the quantum domain.
We were able to include spin into our scheme: 
With the help of a QEP requiring that also spin effects should be 
independent of the mass, we operationally defined a Riemann--Cartan geometry in the quantum domain. 
At least in the non--relativistic domain considered here, the QEP is equivalent to the minimal coupling procedure.

\section{Acknowledgement}

I thank very much Ch.\ Bord\'e and F.W.\ Hehl for discussions and the organisers of the workshop for the invitation. 
Financial support of the CNRS (France) and the WE--Heraeus--Stiftung is acknowledged.


\begin{thebibliography}{99}

\bibitem{ThorneLeeLightman73} K.S.\ Thorne, D.L.\ Lee, A.P.\ Lightman: 
{\it Phys.\ Rev.} {\bf D 7}, 3563 (1973).

\bibitem{Will93} C.M.\ Will: {\it Theory and Experiment in 
Gravitational Physics}, Revised Edition, Cambridge University Press, 
Cambridge 1993. 

\bibitem{AHL92} J.\ Audretsch, F.W.\ Hehl, C.\ L\"ammerzahl {\bf in}: 
Ehlers J., Sch\"afer G.: {\it Relativistic Gravity Research With 
Emphasis on Experiments and Observations}, Springer Lecture Notes in 
Physics {\bf 410}, Springer--Verlag, Berlin 1992, p 368 - 407. 

\bibitem{Rumpf79} Rumpf H. {\bf in}: deSabbata (ed.): {\it International School of Cosmology 
and Gravitation on ``Spin, Torsion, Rotation and Supergravity''}, Plenum, 1979. 

\bibitem{Kleinert78}H.\ Kleinert, this volume.

\bibitem{Pelster78} A.\ Pelster, this volume.

\bibitem{COW75} Colella R., Overhauser A.W., Werner S.A.: {\it Phys. Rev. Lett.} {\bf 34}, 1472 (1975).

\bibitem{Greenberger68} Greenberger D.: {\it Ann. Physics} {\bf 47}, 116 (1968); D.\ Greenberger, A.W.\ Overhauser: {\it Rev. Mod. Phys.} {\bf 51}, 43 (1979). 

\bibitem{Werneretal80} J.--L.\ Staudenmann, S.A.\ Werner, R.\ Colella, A.W.\ Overhauser: {\it Phys. Rev.} {\bf A 21}, 1419 (1980).

\bibitem{La96} C.\ L\"ammerzahl: {\it Gen.\ Rel.\ Grav.} {\bf 28}, 1043 (1996).  

\bibitem{La98} C.\ L\"ammerzahl: {\it Class.\ Quantum Grav.} {\bf 17}, 13 (1998).

\bibitem{AL92} J.\ Audretsch, C.\ L\"ammerzahl {\bf in:} Majer U., 
Schmidt H.-J. (edts.): {\it Semantical Aspects of Space-Time 
Geometry}, BI Verlag, Mannheim 1993.  


\bibitem{La97} C.\ L\"ammerzahl: {\it Phys.Lett.} {\bf A 228}, 223 (1997).

\bibitem{Borde97} Ch.J.\ Bord\'e {\bf in} P.R.\ Berman (ed.): 
{\it Atom Interferometry}, Academic Press 1996, Boston, p.257.

\bibitem{La98a} C.\ L\"ammerzahl, {\bf in}: A.\ Garcia {\it et al.} (eds.) {\it Recent Developments in Gravitation and Mathematical Physics}, 
Science Network Publishing, Konstanz 1998 [online: http://kaluza.physik.uni-konstanz.de/2MS].

\bibitem{AdamowiczTrautmann75} W.\ Adamowicz and A.\ Trautmann: {\it Bull.\ Acad.\ Pol.\ Sci.\ Ser.\ Sci.\ Math.\ Astr.\ \& Phys.} {\bf 23}, 339 (1975). 

\bibitem{Hehletal76} F.W. Hehl, P. von der Heyde, G.D. Kerlick: {\it Rev.\ Mod.\ Phys.} {\bf 48}, 393 (1976). Hehl F.W., Lemke J., Mielke E.W. {\bf in}: Debrus J., Hirshfeld A.C. (edts.): {\it Geometry and Theoretical Phyiscs}, Springer-Verlag, Berlin 1991, p.\ 221.

\bibitem{AudretschLa88} J.\ Audretsch, C.\ L\"ammerzahl: {\it Class. Quantum Grav.} {\bf 5}, 1285 (1988). 

\bibitem{KasevichChu} M.\ Kasevich M., S.\ Chu: {\it Phys. Rev. Lett.} 
{\bf 67} 181 (1991); {\it Appl. Phys.} {\bf B 54}, 321 (1992). 

\bibitem{Trautman66} A.\ Trautman: 
{\it Sov. Phys. Usp.} {\bf 89}, 319 (1966). 

\end{thebibliography}
\end{document}